\documentclass[10pt]{article}
\usepackage[utf8]{inputenc}
\usepackage[T1]{fontenc}
\usepackage{times}
\usepackage{amsmath}
\usepackage{authblk}
\usepackage{natbib}
\usepackage[colorlinks=true,allcolors=blue]{hyperref}
\usepackage{geometry}
\title{Horizontal and Vertical Differentiation: Approaching Endogenous Measurement in Intra-industry Trade}
\author[VIPS-TC, Pitampura, Delhi]{Sourish Dutta}
\date{}

\begin{document}
\maketitle

\begin{abstract}
\noindent Studying intra-industry trade involves theoretical explanations and empirical methods to measure the phenomenon. Indicators have been developed to measure the intensity of intra-industry trade, leading to theoretical models explaining its determinants. It is essential to distinguish between horizontal and vertical differentiation in empirical analyses. The determinants and consequences of intra-industry trade depend on whether the traded products differ in quality. A method for distinguishing between vertical and horizontal differentiation involves comparing exports' unit value to imports for each industry's intra-industry trade. This approach has limitations, leading to the need for an alternative method.
\end{abstract}

\section{Introduction} 
Starting from the end of 1960s, the problem of measuring intra-industry trade lies at the very origin of the literature on intra-industry trade. A series of empirical works describe this phenomenon's intensive \& extensive existence through statistical indicators. Indeed, \cite{greenawayhave} asserts that the literature started with statistical measurement. Among the foundational works on intra-industry trade, the articles of \cite{balassa1965trade, balassa1966tariff} and the book of \cite{grubel1975intra} are especially notable. Because the indicators developed by these authors have acquired a seminal status in the empirical analysis of international trade, these works preceded and triggered the development of different theoretical explanations of intra-industry trade.
	
The theoretical analyses of intra-industry trade explore the phenomenon by adopting disparate assumptions about the nature of returns to scale, the typology of markets, and the differentiation of products subject to this type of trade. \cite{krugman1979increasing}, \cite{lancaster1980intra} and \cite{helpman1981international} explain the rise in intra-industry trade in horizontally differentiated products within a framework of monopolistic competition. \cite{falvey1981commercial} and \cite{shaked1984natural} respectively study the trade of vertically differentiated products in perfectly competitive and oligopolistic markets. Meanwhile, \cite{brander1981intra} models the intra-industry exchange of perfectly homogeneous (and therefore undifferentiated) goods in an oligopolistic market context. 
	
The determinants of intra-industry trade highlighted by these theoretical works are different and depend, in particular, on the type of differentiation taken into account in the various analyses. In monopolistic competition models of \cite{krugman1979increasing}, \cite{lancaster1980intra}, and \cite{helpman1981international}, a necessary condition for the development of intra-industry trade in horizontal differentiation is the existence of economies scale in the production of different varieties. On the other hand, \cite{falvey1981commercial} explains intra-industry trade in a vertical differentiation setting by the difference between the relative factor endowments of countries, where the production of all goods is in the assumption of constant returns to scale. The prediction of \cite{falvey1981commercial} model depends crucially on the following assumption. The production of higher quality differentiated goods requires a greater amount of capital per unit of labour than the production of relatively low-quality ones. However, in the case of measuring intra-industry trade using very disaggregated product classifications, the assumption about the products in that same level of disaggregation having different factorial contents is debatable. In fact, \cite{vona1990intra, vona1991measurement} indicate the similarity between factor contents of products belonging to the same statistical category is very high at the disaggregated product detail levels of international classifications.
	
The theoretical analyses developed during the 1980s help to strengthen the impression that intra-industry trade is a complex phenomenon and multifaceted. Indeed, the Chamberlin-Heckscher-Ohlin (C-H-O) model proposed by \cite{helpman1987market} explains the simultaneous growth of intra-industry trade (in horizontal differentiation) and inter-industry trade between countries. In this framework of economies of scale in the production of differentiated goods, intra-industry trade is the most important if the endowment of relative factors of production of the countries are similar. In contrast, the share of inter-industry trade is an increasing function of the difference between these endowments. 
	
The determinants of intra-industry trade highlighted by \cite{falvey1981commercial} and \cite{helpman1987market} are both different and irreconcilable. Indeed, \cite{falvey1981commercial} predicts a positive relationship between the difference in factor endowments among countries and the rise of intra-industry trade (in vertical differentiation). In contrast, the authors \cite{helpman1987market} show a negative relationship between the difference in factor endowments and the development of intra-industry exchanges (in horizontal differentiation). At the same time, econometric analyses of intra-industry trade were carried out during the 1980s. The works of \cite{balassa1987intra} and \cite{helpman1987imperfect} do not lead to conclusive results of the sign of the existing relationship between the similarity of the factor endowments of trading partners and the share of intra-industry trade in bilateral trade (measured by the index of \cite{grubel1975intra}). \cite{greenaway1997back} express very clearly the lesson that can be drawn from this first wave of econometric analyses aimed at testing intra-industry trade theories:

	\begin{quote}
		"...the determinants for vertical and horizontal intra-industry trade seem to differ. This may explain why different econometric results have led to different conclusions when using total intra-industry trade as the dependent variable. In other words, if the dependent variable in a regression is heterogeneous, it is not surprising that the coefficient for the explanatory variables is somewhat unstable."
	\end{quote}

In other words, econometric models, including an intra-industry trade index total (without distinction between horizontal and vertical differentiation) as an explained variable, are probably poorly specified insofar as the theoretical analysis highlighted different determinants for these two types of trade. Moreover, the theoretical models of \cite{falvey1981commercial} \& \cite{helpman1987market} suggest that the consequences of these two types of intra-industry trade are very different. In particular, adjustment costs (linked to the reallocation of resources from the importing to exporting sectors) generated by the exchanges of vertically differentiated products are greater than those generated through trade in horizontally differentiated products.  
	
All these reasons lead to the conviction that intra-industry trade in horizontal and vertical differentiation constitutes two distinct phenomena regarding determinants and consequences. From the end of the 1980s, this conviction is at the origin of developing methods and statistical indicators to separately measure the extent and evolution of these two types of trade. Thus, the development of research programs concerning intra-industry trade has been a fruitful interaction between theoretical analyses and empirical methods of measuring the importance of this phenomenon. The development of indicators to measure intra-industry trade preceded and triggered the rise of the first theoretical explanations of this phenomenon. These, in turn, have contributed to reviving the empirical debate on the measurement of intra-industry trade by suggesting the need to distinguish, in empirical analyses, intra-industry trade in vertical differentiation from that in horizontal one \citep{balboni2007commerce}.
	
This paper deals with the problem of measuring intra-industry trade. In section 2, it presents two existing approaches to measuring intra-industry trade: the so-called "recovery of trade”, developed by \cite{balassa1966tariff, grubel1975intra} \& the "type of trade” one initiated by \cite{abd1986reexamen, vona1991measurement}. Then this paper presents indicators and empirical methods inspired by these two approaches. Notions of trade recovery \& trade type come from two different definitions of the empirical phenomena they aim to measure. This paper discusses these definitions and the theoretical foundations in section 3. Section 4 presents the existing methods for separating and measuring intra-industry trade in horizontal and vertical differentiation. Section 5 concludes this paper by discussing the division of vertical intra-industry into high and low qualities. 
	
\section{Measurement of Intra-industry Trade}
Two different approaches to the problem of measuring intra-industry trade exist in the discipline of international trade. They are at the origin of several indicators or methods for measuring the level and development of intra-industry trade. \ introduced the Trade Recovery Approach cite{balassa1966tariff} and perfected by \cite{grubel1975intra,grubel1971empirical}. The trade type approach was proposed by \cite{abd1986difference,abd1986reexamen,abd1987hypotheses,abd1991firms} and by \cite{vona1990intra,vona1991measurement} in an independent manner. Subsequently, this section refers to the approaches to the recovery of trade and the type of trade (and the indicators) through the respective acronyms B-G-L (Balassa-Grubel-Lloyd) and A-R-V (Abd-El-Rahman-Vona). Before presenting the details of the indicators developed by these authors to measure intra-industry trade, it is important to note that the B-G-L and A-R-V approaches differ fundamentally in their definition of the phenomenon of intra-industry trade \citep{balboni2007commerce}.

\subsection{The Definitions of Two Approaches}
The B-G-L and A-R-V approach \citep{balboni2007commerce} to measuring intra-industry trade depends on two different conceptions of the phenomenon they aim to understand. This subsection presents the definitions of intra-industry trade (and inter-industry trade) underlying the approaches to the recovery of trade (B-G-L) and type of trade (A-R-V).

\begin{description}
\item[Definition 1:] According to the trade recovery approach, intra-industry trade, for a given industry, is defined as the share of imports and exports (measured in value) perfectly covered (overlapped) by trade flow in the opposite direction. Inter-industry trade is the residual share of flows observed in this industry, i.e. the share of exports or imports not covered by a flow of imports or export in the opposite direction. 

\item[Definition 2:] According to the type of trade approach, for a given industry, if the ratio of the minority flow to the majority flow is not "too low”, then the set flows observed in this industry will be considered intra-industry trade. Otherwise, all flows will be regarded as inter-trade. Here, the minority (majority) flow is the minimum (maximum) important flow between exports and imports in value. The criterion not "too low" is defined more precisely by presenting detailed methods proposed by \cite{abd1987hypotheses,abd1991firms} and \cite{vona1990intra,vona1991measurement} to distinguish between intra-industry and inter-industry trades.
\end{description}

These two definitions imply different methodological choices when measuring intra-industry and inter-industry trades. Measurement methods inspired by the B-G-L approach draw the boundary between intra-industry and inter-industry trades at the interior of each industrial category identifying an industry, while the methods of A-R-V establish this boundary between the different industrial categories. In other words, trade overlap methods separate recorded trade flows by the same industry into two parts: intra-industry trade and inter-industry trade. On the other hand, methods based on the type of trade define the total trade recorded by each industry as an intra-industry trade or an inter-industry trade.

In this regard, a terminological clarification is necessary. Economists using the A-R-V approach prefer to indicate cross-trade \& one-to-one trade as "one-way trade” \& "two-way trade”, respectively \citep{fontagne1997intra,fontagne1998intra}. This choice is consistent with the fact that the recovery methods of exchanges and the methods of type of trade split the total trade in two different ways. Insofar as the intra-industry and inter-industry qualifiers are associated from their introduction by \cite{grubel1971empirical,grubel1975intra}, approaching the recovery of trade, economists using the type of trade approach have proposed two new definitions for the exchanges highlighted in their analyses.

However, the terms "one-to-one trade" and "cross trade" are not used in theoretical works on intra-industry trade, to which the references are made in both analyses utilising the type of trade approach and those using the recovery of trade. For this reason, we have chosen to use the terms "inter-industry trade” and "intra-industry trade” even when we do refer to the methods of the trade type.

\subsection{Indicators of the "Trade Recovery" Approach}
The indicator most used by international economists to measure the intensity of the intra-industry trade between countries is of \cite{grubel1975intra}. The combined form of \cite{grubel1975intra} indicator, which this subsection presents, is called the "synthetic" index of Grubel and Lloyd. The "synthetic" term is to distinguish it from the "simple" Grubel and Lloyd index (referring to trade flows observed in a single industry), which is at the origin. This index was derived from an indicator used by \cite{balassa1966tariff} to measure the share of inter-industry trade in the total trade flow of an industry. 

\subsubsection{Balassa's Indicator}
Indicator of \cite{balassa1966tariff}, denoted $B_i$, measures the relative importance of net trade to the total trade recorded in an industry $i$, defined at a level of given disaggregation of industrial classification. It is calculated by dividing the absolute value of the industry's trade balance $i$ by the sum of exports ($X_i$) and imports ($M_i$) recorded by this industry. The indicator of \cite{balassa1966tariff} can be calculated on a country's trade (or, more generally, of a geographical area) vis-à-vis the rest of the world or a particular country.

\begin{equation}
	\label{equ.1}
	B_i = \frac{\lvert X_i - M_i \rvert}{X_i + M_i}
\end{equation}

The $B_i$ indicator takes values between $0$ and $1$. It is equal to $0$ when the net trade of industry $i$ is zero, i.e. when exports (imports) recorded by the country in question in the industry $i$ perfectly cover the
imports (exports) of the same industry. On the other hand, this indicator equals $1$ when the net trade of industry $i$ coincides with the total trade recorded in this industry, i.e. when one of the two trade flows of this industry (either exports or imports) is equal to zero.

\cite{balassa1966tariff} also develops a variant of the indicator (\ref{equ.1}), in which the balance industry $j$ is not considered in absolute value. 

\begin{equation}
	\label{equ.2}
	B_i = \frac{X_i - M_i}{X_i + M_i}
\end{equation}

Here $B_i$ is an indicator of the trade performance of the concerned country in the industry $i$ to a partner. Naturally, the partner considered can be another country, a geographical area or the rest of the world. In the latter case, indicator $B_i$ will measure the overall trade performance of the country considered in the industry $i$. It varies between $1$ (when the imports are zero) and $-1$ (when exports are zero).

According to \cite{balassa1966tariff}, when the indicator (\ref{equ.2}) is calculated from trade flows relating to industry $i$ of a given country vis-à-vis the rest of the world, it measures the advantage (or the revealed comparative disadvantage) of this country in industry $i$. Using this indicator to understand comparative advantages and trade specialization is open to criticism. The $B_i$ index only considers the trade flows of the industry $i$. And it does not compare the trade in the industry $i$ and the trade flows observed in the others. \cite{balboni2007commerce} concludes that $B_i$ can be considered a trade performance indicator recording only the effects of macroeconomic variables on exports and imports from industry $i$.

\subsubsection{Grubel \& Lloyd's Indicator}
Just like the inter-industry trade indicator (\ref{equ.1}) proposed by \cite{balassa1966tariff}, the index of \cite{grubel1975intra} to measure the intensity of intra-industry trade is based on the definition of these phenomena specific to the approach of recovery of trade flows. Thus, in its simplest version taking into account the trade flows of a single industry, denoted by $i$, the Grubel and Lloyd indicator is calculated as the complement to the unity of Balassa's indicator (\ref{equ.1}).

\begin{equation}
	\label{equ.3}
	GL_i = 1 - B_i = \frac{(X_i + M_i) - \lvert X_i - M_i\rvert}{X_i + M_i} = \frac{2 \min(X_i,M_i)}{X_i + M_i}
\end{equation}

The simple Grubel and Lloyd indicator for industry $i$ (denoted by $GL_i$) is therefore calculated by the
the ratio of overlapped trade flows to total trade recorded by the country considered in industry $i$.

The indicator (\ref{equ.3}), just like the indicator (\ref{equ.1}), can be calculated from the trade flows of a given country with the rest of the world, with a geographical area including various countries, or with a particular country. The third option considers only bilateral trade flows in intra-industry trade indicators to minimize the bias due to the geographical aggregation of the data. This statement is valid for all indices and methods for measuring intra-industry trade.

The indicator (\ref{equ.3}) varies between $0$ and $1$. It takes the maximum value of $1$ when all the trade flows observed in the industry $i$ are intra-industry. It settles at $0$ when all trade in this industry is inter-industry. Indeed, in the first case, the value of exports perfectly covers that of imports. While in the second case, industry $i$ registers a unidirectional trade flow; consequently, the total trade is equal to the trade balance in absolute value.

From indicator (\ref{equ.3}), \cite{grubel1975intra} construct a more sophisticated, which we call the synthetic index of Grubel and Lloyd ($GLS$), allowing us to measure the intensity of intra-industry trade for a grouping (denoted $I$) of $n$ industries (indexed by $i$). The $GLS$ indicator also makes it possible to calculate the intensity of the intra-industry trade on all trade in a country. In this last case, $n$ indicates the total number of industries in the industrial classification retained by the analysis at a given level of disaggregation.

\begin{equation}
	\begin{split}
		\label{equ.4}
		GLS_I &= \frac{\sum_{i=1}^{n}(X_i + M_i) - \sum_{i=1}^{n} \lvert X_i - M_i\rvert}{\sum_{i=1}^{n} (X_i + M_i)} \\
		&= 1 - \frac{\sum_{i=1}^{n} \lvert X_i - M_i\rvert}{\sum_{i=1}^{n} (X_i + M_i)} \\ 
		&= \frac{2 \sum_{i=1}^{n} \min(X_i,M_i)}{\sum_{i=1}^{n} (X_i + M_i)}
	\end{split}
\end{equation}
 
The synthetic index of \cite{grubel1975intra} is the ratio of the sum of trade flows (overlapped) on the total trade of an industry group or a country. Just like the simple $GL_i$ indicator (\ref{equ.3}), this index takes values between $0$ and $1$. It reaches the value of $1$ when all the trade made in industries belonging to the group $I$ is intra-industry in nature, while it settles at $0$ when all the trade flows relating to these industries are of the inter-industry type. 

\subsection{Indicators of the "Type of Trade" Approach}
\cite{abd1986difference,abd1986reexamen,abd1987hypotheses,abd1991firms} and \cite{vona1990intra,vona1991measurement} formulate independently similar criticisms of the indicators belonging to the family of the trade recovery. According to these authors, the measurement of intra-industry trade from these indicators is biased to assume that the flows assimilated to this type of trade are necessarily in balance (overlapped). Given this hypothesis, which is the basis of the definition of intra-industry trade used by \cite{balassa1966tariff} and \cite{grubel1975intra}, the recovery indicators (B-G-L) consider the balanced part of the exchanges carried out within an industry as a type of intra-industry trade and the unbalanced part of these exchanges as a type of inter-industry trade. According to \cite{abd1986difference,abd1986reexamen,abd1987hypotheses,abd1991firms} and \cite{vona1990intra,vona1991measurement}, this dual nature of flows observed within the same industry is a source of confusion, in particular, concerning the identification of the determinants of intra-industry and inter-industry trade. In other words, intra-industry trade cannot be defined as the balanced trade recorded within an industry. This definition implies the attribution of two different natures (inter- and intra-industry) to trade flows observed in the same industry. Conversely, according to these authors, all the trade flows observed within an industry must be considered intra-industry or inter-industry. This principle is the basis of the type of trade (A-R-V) approach. It is affirmed by \cite{vona1991measurement} in the following way: "It is the existence of the simultaneous exchange of very similar goods produced under very similar conditions which constitute intra-industry trade; the existence of an imbalance is irrelevant”.

In this regard, we can see that the semantic choice made by \cite{abd1987hypotheses}, preferring to use the adjectives "one-to-one" and "crossed" instead of the adjectives respective "inter-industry” and "intra-industry”, clearly reflects the desire to apprehend an empirical phenomenon different from the "intra-industry trade” measured by \cite{grubel1975intra}. The two methods based on the principle of type of trade, making it possible to distinguish industries characterized by an intra-industry trade from those with inter-industry trade, are presented in the following. 

\subsubsection{Abd-EI-Rahman's Method}
\cite{abd1987hypotheses,abd1991firms} separates cross trade (intra-industry) from one-way trade (inter-industry) in the following way. By using a very disaggregated product classification to identify the industries empirically, he considers that all trade carried out in a given industry (i.e. the sum of exports and imports relating to this industry) is of unambiguous type when the exchanges observed in this industry satisfy one of the two following conditions:

\begin{itemize}
	\item They are unidirectional (exports or imports are equal to zero);
	\item The minority flow represents less than 10\% of the majority flow (i.e. the smallest flow between exports and imports is less than 10\%, in value of the largest flow).
	\item On the other hand, when (for a given industry) the minority flow is equal to or greater than a tenth of the majority flow, \cite{abd1987hypotheses,abd1991firms} considers that the entire trade carried out in this industry is of cross-type.
\end{itemize}

Thus, \cite{abd1987hypotheses,abd1991firms} sets an arbitrary criterion (the 10\% threshold) allowing to exclude from the field of cross-trade the exchanges carried out in the industries where the flows minority flows are significantly lower than majority flows. The reason for this choice is the conviction that below this threshold, minority flows can have an accidental, not justifying their inclusion in cross-exchanges.

\subsubsection{Vona's Method}
The method proposed by \cite{vona1990intra,vona1991measurement} is very close to that used by \cite{abd1987hypotheses,abd1991firms}. However, it differs from the latter by a more drastic criterion to define the typology of trade flows. \cite{vona1990intra,vona1991measurement} considers only unidirectional flows, observed from disaggregated classification, as inter-industry trade. Therefore, given an industry records both exports and imports of non-zero value, \cite{vona1990intra,vona1991measurement} considers the whole trade observed in this industry (i.e. the sum of its exports and its imports) as an intra-industry trade, even if the minority flow represents a tiny part of the majority flow.

Thus, the definition of intra-industry trade used by \cite{vona1990intra,vona1991measurement} is less restrictive than that inherent in the method of \cite{abd1987hypotheses,abd1991firms}. Nevertheless, both definitions are based on the same principle, according to which the intra-industry or inter-industry concerns all the exchanges carried out in an industry and not only the balanced part of these exchanges. This principle is the basis of the so-called "type of trade” approach to the extent of intra-industry trade. Based on his definition of intra-industry trade, \cite{vona1991measurement} develops an indicator measuring the share of intra-industry trade in the total trade of a grouping (denoted $I$) of $n$ industries (indexed by $i$). We suggest this indicator $VS_I$ (synthetic indicator from \cite{vona1991measurement}).

In the same way as the synthetic index of \cite{grubel1975intra}, the indicator $VS_I$ can be calculated from observed trade in a particular economic sector (composed of $n$ industries) or the total trade flows of a country with a partner. In the latter case, $n$ is the total number of industries identified by the product classification used at a given level of disaggregation. The step before calculating the $VS_I$ indicator consists of the distribution of the $n$ industries into two subsets, according to the typology of trade observed in each industry. This step is used to identify the $m$ industries recording bidirectional trade flows and the $n - m$ industries recording unidirectional trade. Then, the index $VS_I$ is calculated as the ratio between the total trade observed in the $m$ industries recording two-way trade and the total trade relating to the $n$ industries belonging to group $I$.

\begin{equation}
	\label{equ.5}
	VS_I = \frac{\sum_{i=1}^{m} (X_i + M_i)}{\sum_{i=1}^{n} (X_i + M_i)}
\end{equation}

Like the synthetic indicator of \cite{grubel1975intra} (\ref{equ.4}), the indicator (\ref{equ.5}) takes values between $0$ and $1$. It attains the maximum value when all the trade observed in the selected industry group is intra-industry. In this regard, it is important to note that the phenomenon of intra-industry trade, as measured by the indicator of \cite{vona1990intra,vona1991measurement}, is defined in a different way from that highlighted by the indicator of \cite{grubel1975intra}. Indeed, this indicator does not comprise the balanced trade flows (overlapped) observed within each industry. But all the two-way flows are identified at the level of the industries, regardless of their balance.

The indicator (\ref{equ.5}) can also be calculated using the method proposed by \cite{abd1987hypotheses,abd1991firms}, instead of that of \cite{vona1990intra,vona1991measurement}, to distinguish the industries according to the type of trade. In this case, the $m$ industries considered in the index numerator (\ref{equ.5}) are those recording equal minority flows or greater than one-tenth of the majority flows. Thus, when calculated using the method of \cite{abd1987hypotheses,abd1991firms}, the indicator (\ref{equ.5}) records, by construction, lower values than when calculated using the method of \cite{vona1990intra,vona1991measurement}.

\section{Industrial Disaggregation \& Theoretical Foundations}
A recurring problem in empirical analyses of intra-industry trade is the "right choice” of the level of disaggregation of the industrial classification used to define the empirical "industries”. \cite{finger1975trade} initially noted when the industrial categories retained in the empirical analysis to determine the "industries” are not sufficiently disaggregated, they group products characterized by different factor intensities. In this context, a high level of intra-industry trade, measured using the empirical methods described in the previous subsections, constitutes a "statistical illusion”. As cross-flows of products with sufficiently different factor intensities are considered intra-industry trade, ultimately, they indicate inter-industry trade. The problem highlighted by \cite{finger1975trade} is generally defined as the problem of categorical aggregation.

Criticism of \cite{finger1975trade} is addressed in particular to \cite{grubel1975intra}, who use the 3-digit SITC classification to define "industries” in their empirical analysis. This criticism is based on the theoretical definition of the industry-specific to the H-O-S theory of international trade. \cite{finger1975trade} shows that at this level of disaggregation of empirical "industries”, the variability between ratios of the factors used in the production of goods inserted in the same category is more significant than between the proportions of the factors used in the production of goods belonging to different categories. Thus, according to \cite{finger1975trade}, the industry is defined as a group of products characterized by a similar factor intensity at a given level of the relative prices of the generic factors of production. The "industries” retained in the empirical analysis must be consistent with this definition. Otherwise, the results of the analysis will be invalidated. In this regard, we make the following remarks.

The theoretical definition of the industry-specific to the H-O-S model, also retained in the models of \cite{helpman1987market} and \cite{davis1995intra}, is not the only possible theoretical definition of the industry (see Chapter I). For example, this definition is not retained in the theoretical model developed by \cite{falvey1987product}, according to which the same industry includes products characterized by different capital/labour ratios. For these authors, capital is not a generic factor (which can be used indifferently in all industries) but specific to each industry producing differentiated goods. Thus, in their model, the industry is defined as a group of goods whose production requires the implementation of the same factors of production (and not the identical factor intensities).

From this paper's point of view, a preliminary step necessary for any empirical analysis measuring the level and evolution of intra-industry trade consists in specifying the theoretical model retained as the reference explanation of trade flows. The choice of the industrial classification and its level of disaggregation, used to define the empirical "industries”, must be justified by the definition of the industry used in the theoretical reference model.

If the theoretical model retains the H-O-S definition of the industry, we need to look for the best level of disaggregation of the industrial classification. \cite{gullstrand2002does} proceeds in the manner described above when seeking an industrial classification consistent with the theoretical model of \cite{helpman1987market}, retaining the H-O-S definition of the industry. This author asserts that the 6-digit Combined Nomenclature and Harmonized System subheadings include products with similar factorial contents. Thus, he admits that the empirical "industries" corresponding to the 6-digit categories of these classifications respect, in general, the H-O-S definition of the industry. Furthermore, when the H-O-S definition of the industry is used, an excessive industrial disaggregation of the data analyzed can cause biased results in the measurement of inter-industry and intra-industry trade.

From this paper's point of view, the definition of inter- and intra-industry trade adopted in the type of trade approach reflects the juxtaposition of traditional theoretical explanations (Ricardian and H-O-S) of inter-industry trade and explanations of intra-industry trade derived from theoretical analyzes that do not take into account the existence of inter-industry trade. We refer in particular to the theoretical models developed by \cite{krugman1979increasing}, \cite{lancaster1980intra} and \cite{shaked1984natural}, which explain the existence of intra-industry type exchanges between the country, but do not consider the possibility that a trade of an inter-industry nature takes place simultaneously. On the other hand, this definition is inconsistent with the integrated vision of inter- and intra-industry trade specific to the theoretical models that explain the simultaneous development of these two types of trade in a unified analytical framework. See (for example) the theoretical models proposed by \cite{krugman1981intraindustry}, \cite{helpman1987market}, \cite{falvey1987product}, and \cite{davis1995intra}.

The definition of intra-industry trade used in these theoretical models corresponds to that specific to the empirical approach to trade recovery. This definition makes it possible to understand the role of countries' comparative advantages in determining the inter-industrial specialization of their trade, even if this specialization occurs between industries that record two-way trade in differentiated products. Moreover, all these theoretical models suggest that using A-R-V methods to measure the share of intra-industry trade in the bilateral trade between countries carries the risk of underestimating the inter-industrial specialization of the trade when this specialization occurs between industries with differentiated products. Indeed, the A-R-V approach, unlike the B-G-L approach, does not make it possible to highlight these countries' comparative advantages when industries record bidirectional flows.

\section{Measuring Horizontal and Vertical Differentiation}
From the beginning of the 1980s, the first theoretical analysis of intra-industry trade showed that the determinants and consequences of this type of trade are different, depending on whether the traded products differ in quality. When the products are subject to intra-industry trade between two countries with distinct qualities, this trade is vertically differentiated. Otherwise, it is called horizontal differentiation. \cite{abd1986reexamen} proposed a method for distinguishing intra-industry trade between two countries in vertical differentiation from those in horizontal differentiation. This method compares exports' unit value to imports for each industry's intra-industry trade. It considers the trade carried out in this industry as vertical differentiation when the unit value of exports differs significantly from that of imports.

The principle of comparing the unit values of exports and imports, introduced by \cite{abd1986reexamen}, is used in most empirical works about separating intra-industry trade in vertical differentiation from horizontal differentiation. These works also use two methods to measure intra-industry trade flows in vertical and horizontal differentiation: one proposed by \cite{greenaway1994country} and another developed by \cite{fontagne1997intra}. But These two methods measure intra-industry trade in two different ways. The first method is about the trade recovery approach (B-G-L), and the second one retains the type of trade approach (A-R-V). These two approaches are in \cite{balboni2007commerce}. Nevertheless, concerning the separation of exchanges in horizontal and vertical differentiation, the methods of \cite{greenaway1994country} and \cite{fontagne1997intra} apply the same core idea due to \cite{abd1986reexamen}, consisting of comparing the unit value of exports with that of imports.

\subsection{Underlying Assumptions}

The unit value of a trade flow indicates the ratio of its trade value to physical volume. Concerning the physical volume of trade, international trade statistics identify, for a set of categories of products, the number of products exported or imported and, for others, the weight of these products. The method proposed by \cite{abd1986reexamen} assumes that a significant difference observed at the level of a given industrial disaggregation between the unit value of exports and imports reflects a difference in quality between the products exported and those imported. On closer examination, this assumption comprises three nested hypotheses. Those are:
\begin{itemize}
	\item Hypothesis 1: the unit value of exports (imports) observed in an industry reflects the average price of exported (imported) goods belonging to this industry.
	
	\item Hypothesis 2: the prices of goods exported by a given country and belonging to the same industry do not differ significantly. In other words, the dispersion of these prices around their mean is low.
	
	\item Hypothesis 3: the price of a product reflects its quality.
	
\end{itemize}
We now discuss these hypotheses and the issues related to each of them.	

\subsubsection{Discussion of Hypothesis 1}
Among these hypotheses, the first is the most robust from a theoretical and empirical point of view. Nevertheless, we emphasize that the relationship between the unit value of a commercial flow and the average price of the products subject to this flow could not be strictly increasing, particularly in the following case. When, for a given industry, exchanges in volume are counted only in terms of weight, the unit value of the flows relating to this industry corresponds to the average price per ton of the items exchanged and not to their average unit price. In this case, if the prices of exported and imported goods are expressed by unit (and not by weight), a unit value of exports lower than that of imports will not necessarily reflect an average price of exported objects lower than that of products. imported. \cite{greenaway1994country} consider the following example. For some products, greater weight may imply more excellent impact resistance, i.e. longer life. Thus, the unit price of these products increases with their weight, reflecting the better quality of the heaviest objects62. In this context, the products exported by a country may be characterized by an average value per ton lower than or equal to that of the imported products, even if their average price is higher than that of the latter. This case may arise mainly when the country imports lighter and cheaper products (in terms of unit price) than its exports. Thus, when measured in terms of value per ton, the unit value of a commercial flow is not a completely reliable indicator of the average price of the products subject to this flow. 

Comparing unit values per ton of exports and imports can also provide biased information concerning the difference between the average prices of exported and imported products, in the opposite case to that considered by \cite{greenaway1994country}. In some industries (for example, those corresponding to electronic products),  lighter products generally represent higher prices (and quality) than heavier products. In this case, the differences between the per tonne values of exports and imports are much more than the differences between the average prices of exported and imported products. In other words, the difference between the values per ton of exports and imports is, in this case, an "exaggerated" indicator of the difference between the average prices of the products subject to these trade flows.

\subsubsection{Discussion of Hypothesis 2}
Authors who use the unit value to separate intra-industry trade in vertical differentiation from horizontal one consider that all intra-industry trade observed in a given industry is either an exchange of horizontally differentiated products or vertically differentiated products. In the first case, the authors assume de facto that the exported products belonging to the industry considered have a quality similar to that of the imported products. In the second case, the exported products are either higher or lower quality than imported products.

Hypothesis 2 is crucial in comparing the unit values of two trade flows and the relative quality of all the products subject to these flows. This reasoning develops from hypothesis 1, i.e., the unit value of a trade flow reflects the average price of the products covered by this flow and ends with hypothesis 3, according to which the prices of the products reflect their quality. The sequence of these two hypotheses has thus connected hypothesis 2. The relevance of hypothesis 2 depends on the dispersion of prices of the exported or imported products around their average. The greater the standard deviation of these prices, the less their average is significant as an index of the industry's quality of exported (or imported) products.

When the prices of exported (or imported) products belonging to a given industry are very dispersed around their average, it is inaccurate to deduce the average price of exported products and imported products that the former is of a higher or lower quality than the latter. It is also incorrect to conclude that the exported products are similar in quality to the imported products. Indeed, whatever the difference between the average prices of exported and imported products, it is possible that given the significant standard deviation of individual prices, certain exported products (belonging to the industry considered) have prices substantially lower than those of certain imported products. At the same time, other exported products (belonging to the same industry) have higher (or equal) prices than other imported products.

The problems raised by hypothesis 2, unlike those underlying hypotheses 1 and 3, are little debated by economists interested in measuring the relative quality of products subject to international trade. Generally, when the very disaggregated classifications empirically define the industries, the products included in that same industry are relatively homogeneous. Therefore, the assumption is the prices of the different products belonging to the same industry and exported by the same country are not very dispersed around their averages. This assumption makes it possible to assume (when analyzing bilateral export and import flows relating to a given industry) that the unit value of each flow is a significant indicator of prices (and, therefore, of quality) of all the products subject to this flow. The terminology used by \cite{abd1987hypotheses} clearly shows that this author assumes a substantial qualitative homogeneity of the products subject to the same commercial flow. According to this author, a significant difference between the (average) export and import price "suggests that the exported product and the imported product correspond to different qualities"  The expressions "exported product" and "imported product" prove that \cite{abd1987hypotheses} does not take into account the possibility that the same commercial flow (exports or imports) includes varieties of products with prices (and therefore different qualities).

The potential heterogeneity of the products imported by a country in a given industry is even higher if we consider the multilateral trade of a country with different countries in the rest of the world instead of bilateral trade between two countries. Because the prices of products (belonging to a given industry) imported from several countries are probably characterized by a higher standard deviation than those imported from a single country. Thus, the average price level of imported products is generally considered a more reliable indicator of their quality when they come from a single country, i.e., when bilateral trade is regarded as \citep{fontagne1997intra}.

Given the potential heterogeneity of the products included in the same industrial classification, it is likely that the prices of products belonging to the same industry, exported by a given country, are sometimes very dispersed around their averages. This conclusion is more than a simple conjecture as it can be confirmed by analysing the prices of products listed under the same industrial category.

\subsubsection{Discussion of Hypothesis 3}
Hypothesis 3 is generally justified through the following arguments by \cite{greenaway1994country}. On the one hand, in the context of perfect information for economic agents, when two varieties of the same product differ in quality, the higher quality variety is necessarily sold at a higher price. On the other hand, \cite{stiglitz1987causes} shows that even in the context of imperfect information, prices reflect the relative quality of differentiated products. However, economic theory suggests, on the one hand, goods vertically differentiated products are necessarily sold at different prices. Then it teaches, on the other hand, that horizontally differentiated products can also be sold at different prices. In fact, in a monopolistic competitive or differentiated oligopoly market, the prices of differentiated goods of similar quality may differ in equilibrium. 

In a monopolistic competitive market, such as that described by \cite{chamberlin1949theory}, each producer has limited monopoly power, enabling him to set the price of his product above those practised by his competitors without losing all its customers. In a duopolistic market where competitors produce a horizontally differentiated good, it is assumed that the demand functions addressed to the two firms are symmetrical, presenting similar direct and cross-price elasticities and having identical cost functions. In this context, if the two firms simultaneously determine the quantities produced or the prices, the equilibrium prices of the two goods will be identical. On the other hand, if one of the two firms is in a dominant position (which allows it to set its price or quantity by knowing the reaction function of the other firm), the equilibrium prices of the two goods will be different.

We deduce that prices can be considered, at best, as imperfect indicators of product quality.

\subsection{Empirical Approaches}
The method initially proposed by \cite{abd1986reexamen} to separate intra-industry trade in vertical differentiation from those in horizontal differentiation was reformulated and simplified by this same author in later works \citep{abd1987hypotheses,abd1991firms}. The basis of the two versions of this method is on comparing the unit value of exports and imports. These unit values are calculated from bilateral trade flows and listed using detailed industrial classifications. As we anticipated in the previous subsection, for each industry $i$, the unit value of exports (imports), denoted $VUX_i$ ($VUM_i$), is calculated as the ratio between exports (imports) in trade value, denoted $X_i$ ($M_i$) and exports (imports) in volume, denoted $x_i$ ($m_i$).
	
\begin{equation}
	VUX_I = \frac{X_i}{x_i}
\end{equation}
\begin{equation}
	VUM_i =\frac{M_i}{m_i}
\end{equation}

The comparison between the unit values of exports and imports is established by calculating their ratio, which we note $r_i$:
\begin{equation}
	r_i = \frac{VUX_i}{VUM_i}
\end{equation}.

The idea underlying the method of \cite{abd1986reexamen,abd1987hypotheses,abd1991firms} is that a ratio $r_i$ close to $1$ reflects a qualitative similarity of the exported and imported products belonging to the industry $i$. While that ratio $r_i$ moving towards $0$ or $\infty$. testifies to a qualitative difference between the products exported and those imported.

For each industry, $i$, the ratio $r_i$ is confronted with a norm to establish whether the intra-industry trade carried out in this industry must be considered as trade in horizontal or vertical differentiation. The approaches followed by \cite{abd1986reexamen} and \cite{abd1987hypotheses,abd1991firms} are different. We present only the second method as it has established itself in international trade as the reference for separating intra-industry trade into horizontal and vertical differentiation.

After separating the industries characterized by inter-industry (one-to-one) trade from intra-industry (crossed) trade, the method of \cite{abd1987hypotheses,abd1991firms} subsequently distributes the second group of industries into two sets. In the first set, the industries having the difference between the unit value of exports to imports is higher than an arbitrary threshold percentage, set by the author at 15\%, are taken into account. In the second set, industries having the difference between the unit values of exports and imports is less than or equal to 15\%. The trade carried out in those industries belonging to the first set is then considered intra-industry trade in vertical differentiation. At the same time, the industries of the second set are defined as intra-industry trade in horizontal differentiation.

\subsubsection{Arbitrary Threshold}
The criterion proposed by \cite{abd1987hypotheses,abd1991firms} to separate industries carrying out intra-industry trade in horizontal differentiation from those in vertical differentiation is applied in two slightly different ways by \cite{greenaway1994country} and \cite{fontagne1997intra}. After defining an arbitrary threshold $\alpha$ (generally set at 15\% or 25\%), beyond which the difference between the unit value of exports and imports is considered a sign of a difference in quality between the exported and imported products, these authors proceed as described below. 

\cite{greenaway1994country} consider that the products traded in an industry $i$ are horizontally differentiated when the following condition is maintained. 

\begin{equation}
	\label{equ.6}
	1-\alpha \leq r_i \leq 1+\alpha
\end{equation}

Otherwise, they consider that the products traded in industry $i$ are vertically differentiated.

\cite{fontagne1997intra} notice that the right-side term of the condition (\ref{equ.6}) is inconsistent with the left-side one. This inconsistency increases with the value of the arbitrary threshold $\alpha$. Taking condition (\ref{equ.6}) into account implies the possibility that trade in an industry (denoted $i$) for which the $VUX_i / VUM_i$ ratio is equal to the $VUM_j/ VUX_j$ ratio of another industry (denoted $j$), is not considered to be similar (horizontal or vertical) as trade in the industry $j$. It would be logical to attribute a similar nature to trade in both industries.

To illustrate this problem, we assume that the threshold $\alpha$ is set at 15\%. For Industry 1, the unit value of exports equals 1.16, and that of imports equals 1. Whereas, for industry 2, the unit value of imports equals 1.16 and that of exports to 1. In this context, the ratio between the price of export to import for Industry 1 is identical to the ratio between the price of import to export in Industry 2. It would therefore be logical to attribute the exact nature (horizontal or vertical) of the trade flows of these two industries. On the other hand, when condition (\ref{equ.6}) is considered, the intra-industry trade in industry 1 is regarded as a vertical differentiation. In contrast, industry 2 is defined as a trade of horizontal differentiation. Indeed, with the data of this example, the ratios $r_1$ and $r_2$ take the following values: $r_1 = VUX_1 / VUM_1 = 1.16$;  $r_2 = VUX_2 / VUM_2 = 0.86$. Since $r_1 \in (0.85; 1.15)$ and $r_2\in(0.85; 1.15)$, condition (\ref{equ.6}) is satisfied in industry 2 while it is not in industry 1.

Given the inconsistency inherent in the condition (\ref{equ.6}), \cite{fontagne1997intra} consider that the products exchanged in an industry $i$ are horizontally differentiated when the following condition is respected.

\begin{equation}
	\label{equ.7}
	\frac{1}{1+\alpha} \leq r_i \leq 1+\alpha
\end{equation}

When the condition (\ref{equ.7}) is not satisfied, these authors consider that the products traded in industry $i$ are vertically differentiated. When considering condition (\ref{equ.7}) in the context of the numerical example developed above, the intra-industry trade of industry 2 is regarded as an exchange of vertical differentiation, like that of industry 1. Indeed, according to the condition (\ref{equ.7}) (with $\alpha= 15\%$), the interval in which the ratio of unit values $r_i$ must lie to attribute a horizontal nature to trade in industry $i$ is $(0.87; 1.15)$. Consequently, the ratio $r_2 = 0.86$ does not belong to this interval, which implies the assignment of a vertical nature to the intra-industry trade carried out in industry 2.

\subsubsection{Horizontal \& Vertical Differentiation}
We have seen above that \cite{greenaway1994country} and \cite{fontagne1997intra} apply in two slightly different ways the criterion initially suggested by \cite{abd1987hypotheses,abd1991firms} for distinguishing industries performing intra-industry trade in horizontally differentiated products from those developing intra-industry trade in vertically differentiated products.

A more fundamental difference between the approaches proposed by \cite{greenaway1994country} and \cite{fontagne1997intra} concerns how intra-industry trade flows are measured in horizontal and vertical differentiation and their respective shares in total trade. These approaches are based on two different measurements of intra-industry trade, presented and discussed in \cite{balboni2007commerce}: the approach of recovery of trade (B-G-L) and type of trade (A-R-V). \cite{greenaway1994country} adopt the B-G-L approach to the measurement of inter- and intra-industry trade, while \cite{fontagne1997intra} apply the A-R-V approach (and more precisely the method proposed by \cite{abd1987hypotheses,abd1991firms} so to distinguish inter-industry (or one-to-one) trade from intra-industry (or crossed) trade. We also find that two different definitions of intra-industry trade characterise the "B-G-L" and "A-R-V" approaches.

The methods of \cite{greenaway1994country} and \cite{fontagne1997intra} make it possible to measure the importance and the evolution of intra-industry trade in horizontal and vertical differentiation between two countries, presented below. We denote these methods, respectively, GHM and FF. We offer the approaches followed by these authors to measure the relative shares of intra-industry trade in horizontal and vertical differentiation in the total trade observed between two countries in a group (denoted $I$) of $n$ industries, indexed by $i$. Here $IIT_I$, $HIIT_I$, and $VIIT_I$ are the respective shares of intra-industry trade, intra-industry trade in vertical differentiation (Vertical Intra-Industry Trade), and intra-industry trade in horizontal differentiation (Horizontal Intra Industry Trade) in the total trade observed in the group of industries $I$. As the GHM and FF methods break down intra-industry trade into two parts, trade in horizontal and vertical differentiation, the results obtained through these methods always respect the following identity. $ITT_I = HIIT_I + VIIT_I$

\section{Two Parts of Vertical Intra-industry Trade}
	We can add a step to the GHM and FF methods to divide the share of intra-industry trade in vertical differentiation (VIIT) into two parts. The first part corresponds to intra-industry trade flow when a country exports higher quality products than those imported; the second part refers to intra-industry trade for which a country exports lower quality products than imported products. These two sub-parts of the VIIT part are generally noted through the respective acronyms HQVIIT and LQVIIT. HQVIIT (LQVIIT) is the acronym for expressing High Quality (Low Quality) Vertical Intra-Industry Trade. 
	
	First, following the GHM method, we define the set VIIT comprising the industries for which the ratio of unit values $r$ does not satisfy the condition (\ref{equ.1}). Then, we distinguish, within this set, two groups of industries. The first, denoted HQVIIT, includes the industries for which $r_i > 1 + \alpha$. The second, denoted LQVIIT, consists of the industries for which $r_i < 1 - \alpha$. Thus, in the industries of group HQVIIT, the exports of the country considered have unit values higher by at least $\alpha \%$ than the unit values of imports. In the industries belonging to group LQVIIT, the exports have unit values lower by at least $\alpha \%$ than imports. Finally, we calculate the respective shares of intra-industry trade carried out in these two groups of industries in the total trade of the set of industries. By construction, we then have $HQVIIT^{GHM} + LQVIIT^{GHM} = VIIT^{GHM}$, where $VIIT^{GHM}$ is the share of intra-industry trade in vertical differentiation measured by the indicator of \cite{greenaway1994country}. Similarly, it is possible to divide intra-industry trade in vertical differentiation calculated using the FF method of \cite{fontagne1997intra} into two parts, corresponding respectively to trade for which the exported products have a higher quality than the imported products and those for which the exported products are of lower quality than the imported products.
	
\section{Synthesised Discussion}
The methods of GHM or \cite{greenaway1994country} and FF or \cite{fontagne1997intra} are constructed from two different definitions of intra-industry trade. The first method follows the definition of this phenomenon specific to the trade recovery approach. In contrast, the second retains the definition of intra-industry trade regarding the type of trade approach. Consequently, the share of intra-industry trade observed in industries between two countries is measured differently using these two methods. Nevertheless, although they measure intra-industry trade in two different ways,\cite{greenaway1994country} and \cite{fontagne1997intra} use a similar approach to divide this trade into two sub-parts, namely trade in vertical differentiation and horizontal differentiation. This approach, initially suggested by \cite{abd1986reexamen}, involves two basic steps. Firstly, for each industry, $i$, the ratio between the unit value of exports and imports ($r_i$) is comparable to an interval of values fixed arbitrarily. The GHM method retains the interval $(1-\alpha; 1+\alpha)$, while the FF method retains the interval $(1/(1+\alpha); 1+\alpha)$. According to the two methods, the threshold value $\alpha$ is set arbitrarily (generally at 0.15 or 0.25). Secondly, the intra-industry trade flows carried out in the industry $i$ are then considered, in their entirety, either as intra-industry trade in horizontal differentiation or as vertical differentiation, according to whether or not the ratio of unit values $r_i$ is in the fixed interval. This approach has two limitations: the arbitrary nature of threshold $\alpha$ and the attribution of a single quality (horizontal or vertical) to all intra-industry trade flows observed in a given industry.

\section{Problem Identification}
When analysing bilateral trade flows in an industry $i$ over several years, a variation in the ratio of unit values $r_i$ can change the nature assigned to the whole intra-industry trade observed in an industry $i$. When the threshold $\alpha$ is at 0.15, and if the ratio $r_i$ at period $t$ is slightly greater than 1.15 and then a little less than 1.15 at period $t+1$, we can deduce that all of the intra-industry trade observed in industry $i$ is in vertical differentiation at period t. But it is horizontal differentiation at period t + 1. Even if the products exported by each of the two countries in industry $i$ are perfectly homogeneous, the attribution of a horizontal nature to intra-industry trade observed in industry $i$ at period $t$ and vertical nature to this same trade flows at period $t+1$ is open to criticism. Indeed, the ratio between the prices of exported and imported products varies very little between $t$ and $t+1$, and the threshold $\alpha$ determines the "change in nature" of intra-industry trade flows in industry $i$ is arbitrary.
Moreover, if this threshold is at 0.25 instead of 0.15, intra-industry trade flows observed in industry $i$ would be considered intra-industry trade in horizontally differentiated products at both periods of $t$ and $t+1$. Thus the threshold value $\alpha$ influences the result of any empirical analysis relating to the trade flows carried out in a group of unified industries. This influence is more evident when the number of disaggregated industries is small in the aggregate group of industries $I$. In this case, the change in nature of all intra-industry trade carried out in an industry $i$, from one period to another strongly influences the evolution of the respective shares $HIIT_I$ and $VIIT_I$ of intra-industry trade in horizontal and vertical differentiation in the group of industries $I$. For example, using a given threshold may result in abrupt variations in the $HIIT_I$ and $VIIT_I$ shares from one period to another, whereas applying the different threshold values would have highlighted changes in these shares between the same periods. 

The second criticism of GHM and FF methods concerns the principle inherent in assigning a single nature (horizontal or vertical) to all the intra-industry trade flows observed in a given industry. This assignment does not consider the heterogeneity (in terms of price, indicating quality) of the products belonging to a country's each sector. Even though the industries in the empirical analysis are identified with very disaggregated industrial classifications, the exports of a given country in the same industry generally consist of different varieties of products with different prices. In this context, the unit value of exports (imports) recorded for each sector should not be taken as a "true" picture of the quality of all exported (imported) products in that industry but only as an indicator of their average level of quality. Consequently, it is open to criticism as intra-industry trade in products of similar quality and different quality coexist within each industry. 

\section{Solution Proposal}
Therefore, we note the limits of the GHM and FF approaches. Firstly, we assign a unique nature (horizontal or vertical) to the intra-industry trade flows observed in each industry. And secondly, we use an arbitrary threshold, denoted $\alpha$, to determine this nature. This limited approach is open to criticism because there are heterogeneous products in quality to a given industry exported by the same country. Considering this heterogeneity, intra-industry trade in products with similar quality can coexist with different quality within the same industry.

The criticism synthesised above led us to propose an alternative method for separating and measuring intra-industry trade into horizontal and vertical differentiation. The following research will try to show an endogenous approach to make it possible to measure intra-industry trade in horizontal and vertical differentiation (as well as the respective shares of this trade in the total trade flow observed in a group of industries) without using arbitrary thresholds to distinguish these two types. This method will divide the intra-industry trade flow recorded by each industry into two components instead of assigning a single nature to all the trade flows observed in the same industry. We will proceed in the following manner. First, the research will define and construct the indicator of the "verticality" of intra-industry trade. Second, the endogenous measurement of "verticality" and its complementarity will lead to Horizontal and Vertical Differentiation. Third, the research will further divide Vertical Differentiation into two types of value-added, forward and backward.

\bibliographystyle{chicago}
\bibliography{References}
\end{document}